\documentclass[aps,amsfonts,amsmath,nofootinbib,showpacs,byrevtex,preprintnumbers,showkeys,%preprint,%
prd,a4paper]{revtex4} 

\usepackage{graphicx}
\usepackage{bbm,nicefrac}

\newcommand*{\be}{\begin{equation}}
\newcommand*{\ee}{\end{equation}}

\renewcommand*{\vec}[1]{\boldsymbol{\mathrm{#1}}}
\newcommand*{\uvec}[1]{\Hat{\boldsymbol{\mathrm{#1}}}}

\newcommand*{\vx}{\vec{x}}
\newcommand*{\vy}{\vec{y}}
\newcommand*{\vz}{\vec{z}}
\newcommand*{\vk}{\vec{k}}
\newcommand*{\vp}{\vec{p}}
\newcommand*{\vq}{\vec{q}}
\newcommand*{\vl}{\vec{\ell}}
\newcommand*{\uvp}{\uvec{p}}
\newcommand*{\uvq}{\uvec{q}}
\newcommand*{\uvk}{\uvec{k}}

\newcommand*{\braket}[2]{ \langle #1 \vert #2 \rangle }
\newcommand*{\bra}[1]{\langle #1 \rvert \mkern2mu}
\newcommand*{\ket}[1]{\mkern2mu \lvert #1 \rangle}
\newcommand*{\perbra}[2]{ \mkern2mu {}^{^{(#2)}} \mkern-7mu\langle #1 \rvert \mkern2mu }
\newcommand*{\perket}[2]{ \mkern2mu \lvert #1 \rangle \mkern-5mu {}^{^{(#2)}} }
\newcommand*{\perbraket}[4]{ \mkern2mu {}^{^{(#2)}} \mkern-7mu\langle #1 \vert #3 \rangle \mkern-5mu {}^{^{(#4)}} }

\DeclareMathOperator{\Det}{Det}
\DeclareMathOperator{\tr}{tr}

\newcommand*{\comm}[2]{ \bigl[ #1 , #2 \bigr]  }
\newcommand*{\anticomm}[2]{ \bigr\{ #1 , #2 \bigl\}  }

\renewcommand*{\d}[1][]{\mathrm{d}^{#1}}

\newcommand*{\dbar}{\mathrm{d}\mkern-7mu\mathchar'26}
\newcommand*{\deltabar}{\delta\mkern-8mu\mathchar'26}
\newcommand*{\dfr}[2][]{\frac{\mathrm{d}^{#1}#2}{(2\pi)^{#1}}}

\newcommand*{\calD}{\ensuremath{\mathcal{D}}}
\newcommand*{\calJ}{\ensuremath{\mathcal{J}}}
\newcommand*{\calN}{\ensuremath{\mathcal{N}}}
\newcommand*{\calO}{\ensuremath{\mathcal{O}}}

\begin{document}

\title{Perturbation theory in the Hamiltonian approach\\
to Yang--Mills theory in Coulomb gauge}
\date{\today}

\author{Davide R. Campagnari}
\author{Hugo Reinhardt}
\affiliation{Institut f\"ur Theoretische Physik,
Universit\"at T\"ubingen,
Auf der Morgenstelle 14,
D-72076 T\"ubingen,
Germany}
\author{Axel Weber}
\affiliation{Instituto de F\'isica y Matem\'aticas,
Universidad Michoacana de San Nicol\'as de Hidalgo,
Edificio C-3, Ciudad Universitaria, A. Postal 2-82, 58040 Morelia, Michoac\'an, Mexico}

\begin{abstract}
We study the Hamiltonian approach to Yang--Mills theory
in Coulomb gauge in Rayleigh--Schr\"odinger perturbation theory. The static gluon and
ghost propagator as well as the potential between static colour sources
are calculated to one-loop order. Furthermore, the one-loop $\beta$-function is calculated
from both the ghost-gluon vertex and the static potential and found to agree
with the result of covariant perturbation theory.
\end{abstract}

\maketitle

%%%%%%%%%%%%%%%%%%%%%%%%%%%%%%%%%%%%%%%%%%%%%%%%%%%%%%%%%%%
%%%%%%%%%%%%%%%%%%%%%%%%%%%%%%%%%%%%%%%%%%%%%%%%%%%%%%%%%%%

\section{Introduction}

In recent years, there has been a renewed interest in Yang--Mills theory
in Coulomb gauge both in the continuum
\cite{SzcSwa01,FeuRei04,FeuRei04a,ReiFeu05,EppReiSch07,Epp+07,WatRei07a,WatRei07b}
and on the lattice \cite{CucZwa01,LanMoy04,Voi+07,BurQuaRei09}.
This gauge has several advantages over the frequently used
Landau gauge. Among these are: the use of the physical degrees of
freedom (at least in QED) and the explicit emergence of a static
colour charge potential. 
Furthermore, in this gauge the form factor of the ghost propagator 
represents the dielectric function of the Yang--Mills vacuum \cite{Rei08}.
The disadvantage of this gauge is, of course, that it is non-covariant,
which is considered a drawback for perturbation theory, and
renormalisability is yet to be proven.
Recently, there has been much activity in the Hamiltonian approach
to Yang--Mills theory in Coulomb gauge
\cite{SzcSwa01,FeuRei04,EppReiSch07,Epp+07,Rei08,CamRei08,VilAlkSch08,ReiSch08}.
By means of a physically motivated ansatz for the vacuum wave
functional, a variational solution of the Yang--Mills Schr\"odinger
equation has been accomplished \cite{FeuRei04,FeuRei04a,EppReiSch07}. The so-called
gap equation resulting from the minimization of the vacuum energy density
was converted into a set of Dyson--Schwinger equations for the static
gluon, ghost and Coulomb propagators. Due to the particular ansatz
for the vacuum wave functional used, which grasps the essential
infrared physics, the resulting gluon propagator does not yield
the proper ultraviolet (UV) asymptotics known from perturbation theory.
The reason is that the three-gluon vertex does not contribute to the
vacuum energy density for the ansatz of the wave functional considered.

In the present paper, we carry out a thorough
perturbative calculation in the Hamiltonian approach to Yang--Mills theory. We will
use the standard Rayleigh--Schr\"odinger perturbation theory and
carry out the calculations up to and including one-loop order. 
In particular, we will produce the correct one-loop expression for the
perturbative $\beta$-function which has been found in the functional integral
approach \cite{WatRei07a,WatRei07b} before. Because of asymptotic freedom,
we thus have a correct description of the UV regime (to one-loop precision).
With the correct perturbative results at hand, we show
how the variational approach has to be modified to yield
the correct UV asymptotics. 
Furthermore, the perturbative calculations performed
in the present paper can be extended to calculate systematic corrections to the
non-perturbative variational vacuum solution and to construct a complete basis
for the Hilbert space of Yang--Mills theory, which is required for the
calculation of the partition function or free energy. In a forthcoming paper,
the latter will be minimized to achieve a description of the deconfinement phase
transition at finite temperatures in the Hamiltonian approach.

%%%%%%%%%%%%%%%%%%%%%%%%%%%%%%%%%%%%%%%%%%%%%%%%%%%%%%%%%%%
%%%%%%%%%%%%%%%%%%%%%%%%%%%%%%%%%%%%%%%%%%%%%%%%%%%%%%%%%%%

\section{\label{sec:ham}Perturbative expansion of the Yang--Mills Hamiltonian}

\subsection{The Yang--Mills Hamiltonian in Coulomb gauge}

In the absence of matter fields, the Yang--Mills Hamiltonian in Coulomb gauge
reads \cite{ChrLee80}
\begin{equation}\label{G1}
\begin{split}
{H}_\mathrm{YM} = \int \d[d]x \: &\left[ \frac12 \, \calJ^{-1} \,
\Pi^a_i(\vx) \, \calJ \, {\Pi}^a_i(\vx)  +
\frac14 \: F_{ij}^a(\vx) \, F_{ij}^a(\vx) \right] + \\
&+ \frac{g^2}{2}  \int \d[d]x \, \d[d]y \:
\hat{A}^{ac}_i(\vx) \, \calJ^{-1} \, \Pi^c_i(\vx) \, \calJ \, F^{ab}[A](\vx,\vy) 
\hat{A}^{bd}_j(\vy) {\Pi}^{d}_j(\vy) \, ,
\end{split}
\end{equation}
where $\vx$ is a vector in $d$ space dimensions, $A_i^a$ is the transverse gauge
field operator, $\Pi_i^a$ is the transverse momentum
operator satisfying the canonical commutation relations
\begin{equation}\label{ccr}
\comm{A_i^a(\vx)}{\Pi_j^b(\vy)} = i \, \delta^{ab} \,
t_{ij}(\vx) \, \delta(\vx-\vy) \, , \quad
t_{ij}(\vx)= \delta_{ij} - \partial_i \partial_j/\partial^2 \, ,
\end{equation}
and
\begin{equation}
F_{ij}^a = \partial_i A_j^a - \partial_j A_i^a + g \, f^{abc} A_i^b A_j^c
\end{equation}
is the field strength tensor ($g$ is the coupling constant, $f^{abc}$ are
the structure constants of the $\mathfrak{su}(N_c)$ algebra and
$\Hat{A}^{ac}=f^{abc} A^b$ is the gauge field operator in the
adjoint colour representation). Furthermore,
\be
\label{G4}
\calJ[A] = \Det (- \hat{D} \cdot \partial) / \Det (- \partial^2) \, , \quad
\Hat{D}_i^{ab} = \delta^{ab} \, \partial_i + g \Hat{A}_i^{ab}
\ee
is the Faddeev--Popov determinant of Coulomb gauge, and
\begin{equation}
\label{coulomb.operator}
F^{ab}[A](\vx,\vy) = \left[ (-\hat{D} \cdot \partial)^{-1} \: (-\partial^2) 
\: (-\hat{D} \cdot \partial)^{-1} \right]^{ab}_{\vx,\vy}
\end{equation}
the so-called Coulomb kernel. 

The Coulomb gauge $\partial_i A_i^a = 0$ is implemented in the scalar product
of the Hilbert space of Yang--Mills wave functionals by the Faddeev--Popov
method, yielding for the matrix elements of observables $O[A,\Pi]$
\be
\label{G3}
\bra{\psi_1} O \ket{\psi_2} = \int \calD A \: \calJ[A] \, \psi^*_1[A] \, O[A,\Pi] \, \psi_2[A] \: ,
\ee
where the integration is over the transverse gauge fields only.

It is convenient to remove
the Jacobian $\calJ$ from the integration measure by defining
\be
\label{G5} \psi [A] = \calJ^{-\nicefrac{1}{2}} \: \widetilde{\psi} [A] \quad , \quad
\widetilde{O} = \calJ^{\nicefrac{1}{2}} O \calJ^{-\nicefrac{1}{2}}
\ee
The transformed Hamiltonian $\widetilde{H}$ is then obtained from Eq.\ (\ref{G1})
by replacing the momentum operator $\Pi$ by the transformed one
\be
\label{G6}
\widetilde{\Pi}_i^a(\vx)
= \calJ^{\nicefrac{1}{2}} \Pi_i^a(\vx) \calJ^{-\nicefrac{1}{2}} 
= \Pi_i^a(\vx) + \frac{i}{2} \frac{\delta \ln \calJ}{\delta A_i^a(\vx)} \: .
\ee
This yields
\begin{equation}\label{h-YM}
\begin{split}
\widetilde{H}_\mathrm{YM} = \int \d[d]x \: &\left[ \frac12 
\widetilde{\Pi}^{a\dagger}_i(\vx) \, \widetilde{\Pi}^a_i(\vx)  +
\frac14 F_{ij}^a(\vx) \, F_{ij}^a(\vx) \right] + \\
&+ \frac{g^2}{2}  \int \d[d]x \, \d[d]y \:
\hat{A}^{ac}_i(\vx) \widetilde{\Pi}^{c\dagger}_i(\vx) F^{ab}[A](\vx,\vy) 
\hat{A}^{bd}_j(\vy) \widetilde{\Pi}^{d}_j(\vy) \, .
\end{split}
\end{equation}
In the following, Dirac's bra-ket notation will refer to the
transformed space where no Faddeev--Popov determinant occurs
in the functional integration measure.

For perturbation theory, it will be convenient to expand the gauge
field in Fourier modes. We use the following conventions
\be
\label{G7}
A_i^a(\vx) = \int \dfr[d]{p} \: e^{i \vp \cdot \vx} \, A_i^a (\vp) \: ,
\quad \Pi_i^a (\vx) =  \int \dfr[d]{p} \: e^{i \vp \cdot \vx} \, \Pi_i^a (\vp)  \, ,
\ee
where the %$p_i = 2 \pi n_i/V^{1/d}$ are the Matsubara frequencies and the
transformed fields satisfy the canonical commutations relations
\be
\label{G9}
\comm{A^a_i(\vk)}{\Pi^b_j(\vp)} = i \delta^{a b} t_{ij}(\vk) \,
(2\pi)^d \, \delta(\vk+\vp) \: ,
\ee
and where $t_{i j} (\vk) = \delta_{i j} - k_i k_j/\vk^2$ is the transverse projector
in momentum space. To simplify the notation, we introduce the following shortcuts
\be\label{G10}
\dbar{p} \equiv \dfr[d]{p} \: , \quad
\deltabar(\vp) \equiv (2\pi)^d \delta(\vp) \, .
\ee

%%%%%%%%%%%%%%%%%%%%%%%%%%%%%%%%%%%%%%%%%%%%%%%%%%%%%%%%%%%

\subsection{Expansion of the Hamiltonian}

Expanding the Hamiltonian $\widetilde{H}_\mathrm{YM}$ \eqref{h-YM} in powers of the coupling constant $g$,
thereby using $\calJ = 1 + \calO (g^2)$ and
\be
\label{G11}
\widetilde{\Pi}^\dagger \widetilde{\Pi} = \Pi^2 - \calJ^{-\nicefrac{1}{2}} \comm{\Pi}{\comm{\Pi}{\calJ^{\nicefrac{1}{2}}}} \; ,
\ee
we obtain
\be
\label{G12}
\widetilde{H} = H_0 + g \widetilde{H}_1 + g^2 \widetilde{H}_2 + \calO (g^3) \: ,
\ee
where the unperturbed Hamiltonian
\be\label{h0}
H_0 = \frac12 \int \d[d]x \left[ \big( \Pi_i^a(\vx) \big)^2 - A_i^a(\vx) \, \partial^2 A_i^a(\vx) \right]
\ee
is the Hamiltonian of QED except for the extra colour index of the gauge field. 
The first order term $g \, \widetilde{H}_1$ arises from the expansion of the magnetic energy $\int \d[d]
x F^2_{i j}$ and is given by the three-gluon vertex
\be\label{ham-1}
\widetilde{H}_1 = \frac{i}{3!} \: f^{a_1a_2a_3} \int \dbar{k_1} \, \dbar{k_2} \, \dbar{k_3} \, \deltabar(\vk_1+\vk_2+\vk_3) \:
T(1,2,3) \, A(1) \, A(2) \, A(3) \: .
\ee
Here we have introduced the shorthand notation
$A (1) \equiv A^{a_1}_{i_1}(\vk_1)$ and  $T(1,2,3)$ carries the
(totally antisymmetric) Lorentz structure of the three-gluon vertex,
\be\label{3.gluon.vertex}
T(1,2,3):= t_{i_1j}(\vk_1) \, t_{i_2l}(\vk_2) \, t_{i_3m}(\vk_3)
\big[ \delta_{jl} (k_2 - k_1)_m + \delta_{lm} (k_3 - k_2)_j + \delta_{jm} (k_1 - k_3)_l \big] .
\ee
Finally, the second-order term
\begin{equation}\label{ham-2}
\begin{split}
\widetilde{H}_2 =  \frac{1}{2} \, & f^{aa_1a_2}f^{aa_3a_4} 
 \int \dbar{k_1} \ldots \dbar{k_4} \, \deltabar(\vk_1+\vk_2+\vk_3+\vk_4) \\
& \times \bigg[ \frac{\delta_{i_1i_3} \, \delta_{i_2i_4}}{2} \, A(1) \, A(2) \, A(3) \, A(4)
+ \frac{\delta_{i_1i_2} \, \delta_{i_3i_4}}{(\vk_1+\vk_2)^2} \, A(1) \, \Pi(2) \, A(3) \, \Pi(4) \bigg] + C .
\end{split}
\end{equation}
contains besides the usual four-gluon vertex (first term in the bracket) also a contribution
from the Coulomb term (second term in the bracket) arising from the expansion of the Coulomb
kernel \eqref{coulomb.operator}. Note that the Coulomb term is already
$\calO(g^2)$, see Eq.~\eqref{h-YM}, so that to the order considered we can replace the
Coulomb kernel \eqref{coulomb.operator} simply by its bare form
$(-\partial^2)^{-1}$. The last term in Eq.~(\ref{ham-2}) is an irrelevant constant arising
from the expansion of the second term in Eq.~(\ref{G11}). Since such a constant
does not influence the wave functional, we will skip it in the following.

Since we will use dimensional regularisation, in order to preserve
the dimension of the dressing functions we will replace
\be
\label{G17}
g \to g \mu^{(3 - d)/2} ,
\ee
with $\mu$ being an arbitrary mass scale.

%%%%%%%%%%%%%%%%%%%%%%%%%%%%%%%%%%%%%%%%%%%%%%%%%%%%%%%%%%%

\subsection{\protect\label{sec:basis}The unperturbed basis}

The perturbative vacuum state $\psi_0 [A] = \langle A | 0 \rangle$ is given by
\be\label{pert.vacuum}
\langle {A}|{0} \rangle = \calN \exp \left\{ - \frac{1}{2} \int \dbar{k} \: A_i^a(\vk) \, t_{ij}(\vk) \, |\vk| \, A_j^a(-\vk) \right\} .
\ee
Up to the colour index of the gauge field, this is precisely the exact vacuum
wave functional for QED without fermions. This state is the lowest energy eigenstate of $H_0$
(\ref{h0}) with energy
\be
\label{G19}
E_0 = (N_c^2-1) \frac{d-1}{2} \int \dbar{k} \: |\vk| \: .
\ee
To restrict the Coulomb gauge field to its transverse degrees of freedom, it is
convenient to introduce the eigenvectors of the transverse projector 
$t_{i j} (\vk)$ in $d$ spatial dimensions with eigenvalue $1$
\be
\label{G20}
t_{i j} (\vk) c^\sigma_j (\vk) = c^\sigma_i (\vk) \: ,
\ee
where $\sigma$ labels the $d - 1$ different eigenvectors. From $k_i t_{ij} (\vk) = 0$
it follows immediately that these eigenvectors are orthogonal
to $\vk$
\be
\label{G21}
k_i c^\sigma_i (\vk) = 0 \: .
\ee
Assuming the normalisation 
\be\label{quasiparticle-3}
{c_i^\sigma}^*(\vk) \, c_i^\tau(\vk) = \delta_{\sigma\tau} \: ,
\ee
these vectors satisfy the ``completeness'' relation in the transverse
subspace
\be
\label{G23}
c^\sigma_i (\vk) c^{\sigma^*}_j (\vk) = t_{i j} (\vk) \: .
\ee
Since the transverse projector $t_{i j} (\vk)$ is a real symmetric matrix,
it can be diagonalised by an orthogonal transformation and hence the
eigenvectors $c^\sigma_i (\vk)$ can be chosen real, which we will assume
below. Let us mention, however, that in $d = 3$ the $c^\sigma_i (\vk)$ are
usually chosen as the circular polarisation vectors, which are complex.

The transverse components of the gauge field and their momenta are then given by
\be
\label{G24}
A^a_\sigma (\vk) := c^\sigma_i (\vk) A^a_i (\vk) \: , \quad \Pi^a_\sigma
(\vk) := c^\sigma_i (\vk) \Pi^a_i (\vk) \:  .
\ee
In view of Eq.~(\ref{G9}) they satisfy the commutation relation
\be
\label{G25}
\comm{A^a_\sigma(\vk)}{\Pi^b_\tau(\vp)} = i \delta_{\sigma \tau} \delta^{a b} \deltabar(\vk+\vp) \: .
\ee
In the transverse components of the gauge field, the perturbative vacuum state
(\ref{pert.vacuum}) reads
\be
\label{G26}
\braket{A}{0} = \calN \exp \left\{ - \frac{1}{2} \int \dbar{k} \:
A^a_\sigma(\vk) \, |\vk| \, A^a_\sigma(-\vk) \right\} \: .
\ee
This state is annihilated by the operator 
\be
\label{G27}
a^a_\sigma (\vk) = \sqrt{\frac{|\vk|}{2}} \left[A^a_\sigma (\vk) +
\frac{i}{| \vk|} \Pi^a_\sigma (\vk) \right] , \quad a^a_\sigma (\vk) | 0 \rangle
= 0 \: ,
\ee
which together with its hermitean conjugate
\be
\label{G28}
a_{\sigma}^{a\dagger}(\vk) = 
\sqrt{\frac{|\vk|}{2}} \left[ A^a_\sigma(-\vk) - \frac{i}{|\vk|} \, 
\Pi^a_\sigma(-\vk) \right] ,
\ee
fulfills the usual Bose commutation relation
\be
\label{G29}
\comm{a^a_\sigma (\vk)}{a^{b\dagger}_\tau (\vp)} = \delta^{a b} \delta_{\sigma\tau}
\deltabar(\vk-\vp) \: .
\ee
The unperturbed Hamiltonian Eq.~(\ref{h0}) is diagonalised by the
transformations \eqref{G27}, \eqref{G28}
\be
\label{G30}
H_0 = E_0 + \int \dbar k \: | \vk | \, a^{a\dagger}_\sigma (\vk) \, a^a_\sigma (\vk) \: ,
\ee
and accordingly the eigenfunctions of $H_0$ are the multiple gluon states
defined by
\be
\label{G31}
\lvert g_1 g_2 \dots g_N \rangle = a^{a_1 \dagger}_{\sigma_1}(\vk_1) \,
a^{a_2 \dagger}_{\sigma_2}(\vk_2) \dots a^{a_N \dagger}_{\sigma_N}(\vk_N) \, | 0 \rangle
\ee
with energies
\be
\label{G32}
E_0 + \sum^N_{i = 1} | \vk_i | \: .
\ee
There are two ways to proceed now: Given the unperturbed basis (\ref{G31}), one
can express the gauge field $A$ and its momentum operator $\Pi$ in the
perturbations, Eqs.~\eqref{ham-1} and \eqref{ham-2}, in terms of
the creation and annihilation operators $a^\dagger_\sigma ,
a_\sigma$. Alternatively, since the perturbation $\widetilde{H}_{1,2}$
is expressed in terms of the gauge field and its momentum, one may wish to
express the unperturbed basis states through
$A$ and $\Pi$ using Eq.~(\ref{G28}). This yields
\be\label{basis}
\ket{g} = \sqrt{2|\vk|} \; A_\sigma^a(-\vk) \ket{0} \, .
\ee
for a one-particle state, and
\be
\label{G34}
\ket{g_1,g_2} = 
\left[ 2 \sqrt{|\vk_1| \, |\vk_2|} \; A^{a_1}_{\sigma_1}(-\vk_1) \, A^{a_2}_{\sigma_2}(-\vk_2) 
- \delta^{a_1 a_2} \, \delta_{\sigma_1 \sigma_2} \deltabar(\vk_1+\vk_2) \right] \ket{0} \, ,
\ee
for a two-particle state, where the second term on the right hand side of the last equation arises from
the canonical commutation relations. States with more gluons have similar
additional contraction terms. These contact terms ensure
that the free $n$-gluon states are orthogonal to each other and to the unperturbed vacuum.
It turns out that the contact terms simply eliminate from the matrix
elements of observables the contractions of gauge field operators stemming
exclusively from the wave functionals. We can therefore use the simplified
representation
\be
\label{G35}
\ket{g_1, \ldots , g_n} = \left[ \prod_{i=1}^n \sqrt{2 |\vk_i|} \; A_{\sigma_i}^{a_i}(-\vk_i) \right] \ket{0} \, ,
\ee
with the additional calculational rule that in the evaluation of matrix elements
of the form $\bra{0} O[A,\Pi]\ket{g_1, \ldots , g_n}$, the $A$ fields in the
states Eq.~\eqref{G35} must not be contracted with each other but only with
$A$, $\Pi$ occurring in the observable $O[A,\Pi]$ .

%%%%%%%%%%%%%%%%%%%%%%%%%%%%%%%%%%%%%%%%%%%%%%%%%%%%%%%%%%%

\subsection{Expansion of the vacuum wave functional}

In Rayleigh--Schr\"odinger perturbation theory, the leading order (in the coupling
constant) corrections to the vacuum wave functional are given by
\begin{subequations}\label{pert.corrections}
\begin{align}
\perket{0}{1} &= - \sum_n \frac{1}{n!} \: \frac{\bra{g_1 \ldots g_n} \widetilde{H}_1 \ket{0}}{|\vk_1| + \ldots + |\vk_n|}
\: \ket{g_1 \ldots g_n} \, , \label{1.order.vacuum} \\
\perket{0}{2} &= - \sum_n \frac{1}{n!} \: \frac{1}{|\vk_1| + \ldots + |\vk_n|} \left[ \bra{g_1 \ldots g_n} \widetilde{H}_2 \ket{0}
+ \bra{g_1 \ldots g_n} \widetilde{H}_1 \perket{0}{1} \right]
\ket{g_1 \ldots g_n} \, , \label{2.order.vacuum}
\end{align}
\end{subequations}
where $\widetilde{H}_1$ and $\widetilde{H}_2$ are defined in Eqs.~(\ref{ham-1}),
(\ref{ham-2}), and the factors $1/n!$ avoid multiple counting due to
identical gluons (summation over colour and polarisation indices and
integration over the momenta is implicit).

It is now straightforward to calculate the perturbative corrections to the
vacuum wave functional. Since $\tilde{H}_1$ contains three field operators and is
antisymmetric in both colour and Lorentz indices, in view of Eq.~(\ref{G35}) it is
clear that only three-gluon states will contribute to
$\perket{0}{1}$, Eq.~(\ref{1.order.vacuum}), yielding
\be\label{1st.order.vacuum}
\phantom{}\perket{0}{1} = \frac{i}{3!} \: f^{a_1a_2a_3} \int \dbar{k_1} \, \dbar{k_2} \, \dbar{k_3} \: 
\frac{\deltabar(\vk_1+\vk_2+\vk_3)}{|\vk_1| + |\vk_2| +|\vk_3|} \: T(1,2,3) \: A(-1) \, A(-2) \, A(-3) \ket{0} \: .
\ee
As a consequence, the second term
in $\perket{0}{2}$, Eq.~(\ref{2.order.vacuum}), receives contributions from up to
six-gluon states. Furthermore, since $\widetilde{H}_2$ contains terms with up
to four field operators, the first term in $\perket{0}{2}$
will receive contributions from two- and four-gluon
states. However, it turns out that up to order $g^2$ only
two-gluon states contribute to the static propagators, so that
\be\label{2nd.order.vacuum}
\phantom{}\perket{0}{2} = - N_c \: \frac{\delta^{a_1a_2}}{8} \int \dbar{k_1} \, \dbar{k_2}
\frac{\deltabar(\vk_1+\vk_2)}{|\vk_1|} \left[ F_B(1,2) + F_C(1,2) + F_1(1,2) \right] A(-1) \, A(-2) \ket{0} \: ,
\ee
where we have introduced the abbreviations
\begin{subequations}
\begin{align}
F_B(1,2) &= \frac12 \int \dbar{q} \: \frac{(d-1) t_{i_1i_2}(\vk_1) - t_{i_1i_2}(\vq)}{|\vq|} \: , \\
F_C(1,2) &= \int \dbar{q} \: \frac{t_{i_1i_2}(\vq)}{(\vk_1-\vq)^2} \left[ |\vq| - \frac{\vk_1^2}{|\vq|} \right]  \: , \\
F_1(1,2) &= -\frac12 \int \dbar{k_3} \, \dbar{k_4} \, \frac{T(-1,3,4) \, T(2,3,4)}{|\vk_1| + |\vk_3| +|\vk_4|}
\frac{\deltabar(\vk_2+\vk_3+\vk_4)}{|\vk_3| \, |\vk_4|} \: .
\end{align}
\end{subequations}
Here, $F_B (1, 2)$ arises from the four-gluon vertex, $F_C (1, 2)$ from the
Coulomb term, and $F_1 (1, 2)$ from the second-order contribution of the three-gluon
vertex, see the last term in Eq.~(\ref{2.order.vacuum}). We have also used
the shorthand notation $A (- 1) = A^{a_1}_{i_1} (- \vk_1)$, etc.

In the Rayleigh-Schr\"odinger perturbation theory used in
Eqs.~(\ref{pert.corrections}), the correction to a wave
function of a given order is chosen to be orthogonal to
the unperturbed wave function,
\be
\label{G43}
\langle 0 \! \perket{0}{i} = 0 \: , \: i \geq 1 \: .
\ee
As a consequence, the wave functions obtained in a given order of perturbation
theory are not properly normalised. We are interested in the vacuum wave
functional up to second order
\be
\label{G44}
\ket{\Omega} = \calN_\Omega \, \left[ \ket{0} + g \perket{0}{1} + g^2 \perket{0}{2} + \calO(g^3) \right] .
\ee
Calculating the normalisation constant $\calN_\Omega$ up to order $g^2$ 
making use of Eq.~(\ref{G43}), we find for the properly normalised vacuum wave
functional up to this order
\be\label{full.vacuum}
\ket{\Omega} = \left[ 1- \frac{g^2}{2} \perbraket{0}{1}{0}{1} \right] \! \ket{0} +
g \perket{0}{1} + g^2 \perket{0}{2} + \calO(g^3) \: .
\ee
In the following sections we will use this perturbative expansion of the vacuum
wave functional to calculate various static propagators.

%%%%%%%%%%%%%%%%%%%%%%%%%%%%%%%%%%%%%%%%%%%%%%%%%%%%%%%%%%%
%%%%%%%%%%%%%%%%%%%%%%%%%%%%%%%%%%%%%%%%%%%%%%%%%%%%%%%%%%%

\section{\label{sec:ghost}Ghost propagator}

The Green's function (or inverse) $G[A]$ of the Faddeev--Popov
operator is defined by
\be\label{ghost.green.function}
- \Hat{D}_i^{ab}[A](\vx) \partial_i^x \, G^{bc}[A](\vx,\vy) = \delta^{ac} \, \delta(\vx-\vy) \: .
\ee
Expanding $G[A]$ in a power series in the coupling constant
\be\label{ghost.pert.expansion}
G^{ab}[A](\vx,\vy) = \sum_{n=0} g^n \, G_n^{ab}[A](\vx,\vy)
\ee
we get from (\ref{ghost.green.function}) the recursion relation
\begin{subequations}
\be\label{ghost.recursion}
- \partial^2_x \, G^{ab}_{n+1}[A](\vx,\vy) = \Hat{A}_i^{ac}(\vx) \partial_i^x \, G^{cb}_n[A](\vx,\vy) \, , \quad n\geq0
\ee
together with the initial condition
\be\label{ghost.initial.condition}
- \partial^2_x \, G^{ab}_0[A](\vx,\vy) = \delta^{ab} \, \delta(\vx-\vy) \: ,
\ee
\end{subequations}
which defines the unperturbed static ghost propagator $G^{ab}_0[A](\vx,\vy)$ as the
Green's function of the Laplacian, which in momentum space reads
\be\label{ghost.perturbative}
G_0^{ab}[A](\vk) = \delta^{ab} \, G_0(\vk) = \frac{\delta^{ab}}{\vk^2} \: .
\ee
It is diagonal in colour space and independent of the gauge field.
With this  property, it follows from  Eq.~\eqref{ghost.recursion}
that each term $G_n[A]$ contains a product of $n$
gauge field operators.

The ghost propagator is defined as the
expectation value of the inverse Faddeev--Popov operator
in the vacuum state $|\Omega\rangle$,
\be\label{ghost.propagator}
G(\vx,\vy) = \bra{\Omega} G[A](\vx,\vy) \ket{\Omega} \, .
\ee
Since $G [A]$ does not depend on the momentum operator $\Pi$, it does not change
under the transformation to the ``radial'' Hilbert space, \textit{i.e.} $\tilde{G} [A] =
G [A]$, see Eq.~(\ref{G5}).
Furthermore, contrary to the Faddeev--Popov operator, the vacuum expectation value of its
inverse, the static ghost propagator, is translationally invariant.

Inserting the expansions \eqref{ghost.pert.expansion} for $G[A]$
and \eqref{full.vacuum} for $\ket{\Omega}$ into Eq.~\eqref{ghost.propagator}
it is possible to show that many terms vanish or cancel, so that
the ghost propagator reduces to
\be\label{eq53}
G(\vx,\vy) = G_0(\vx,\vy) + g^2 \bra{0} G_2[A](\vx,\vy) \ket{0} + \calO(g^3)\, .
\ee
The second-order term in Eq.~\eqref{eq53} can be evaluated by means
of the recursion relation \eqref{ghost.recursion}, yielding
\be\label{2.order.ghost}
G_2^{ab}[A](\vx,\vy) =
\left[ G_0 \: (\Hat{A} \cdot \partial) \: G_0 \: (\Hat{A} \cdot \partial) \: G_0\right]^{ab}_{\vx,\vy} \: ,
\quad G_0=(-\partial^2)^{-1}
\ee

The vacuum expectation value of Eq.~\eqref{2.order.ghost} can be expressed 
through the bare static gluon propagator, which in view of
Eq.~\eqref{pert.vacuum} reads
\be\label{tree.level.gluon}
D_0 (1, 2) = \bra{0} A(1) \, A(2) \ket{0} 
 = \delta^{a_1 a_2} \deltabar(\vk_1+\vk_2) \:
\frac{t_{i_1 i_2} (\vk_1)}{2 |\vk_1|} \: .
\ee
Then for the ghost form factor $D_c(\vk)$ defined by
\be\label{ghostformfactor}
G(\vx,\vy) =: \int \dbar{k} \:  e^{i\vk\cdot(\vx-\vy)} \: \frac{D_c(\vk)}{\vk^2} \: ,
\ee
we get the following expression at one-loop order
\be
\label{875-37}
D_c(\vk) = 1 + g^2 \, \mu^{3-d} \: \frac{N_c}{2\vk^2} \int \dbar{q} \:
\frac{k_i \, k_j \, t_{ij}(\vq)}{(\vk-\vq)^2 \; |\vq|} \: .
\ee
The integral (\ref{875-37}) is standard
and can be evaluated in dimensional regularisation with
$d=3-2\varepsilon$ in the usual way, yielding
\be\label{ghost-dimreg}
D_c(\vk) = 1 + g^2 \; \frac{N_c}{(4\pi)^{2-\varepsilon}} 
\left[ \frac{4}{3} \left( \frac{1}{\varepsilon} - \ln\frac{\vk^2}{\mu^2} - \gamma\right) - \frac{8}{3} \; \ln2 + \frac{28}{9}
+ \calO(\varepsilon) \right] .
\ee

%%%%%%%%%%%%%%%%%%%%%%%%%%%%%%%%%%%%%%%%%%%%%%%%%%%%%%%%%%%
%%%%%%%%%%%%%%%%%%%%%%%%%%%%%%%%%%%%%%%%%%%%%%%%%%%%%%%%%%%

\section{Gluon propagator}

\subsection{\protect\label{sec:gluon}Gluon propagator in the Hamiltonian approach}

The full static gluon propagator is defined in momentum space by
\be
\deltabar(\vk+\vp) \: D_{ij}^{ab}(\vk) = \bra{\Omega} A_i^a(\vk) \, A_j^b(\vp) \ket{\Omega} \: .
\ee
With the expansion \eqref{full.vacuum} for the
vacuum functional, the nonvanishing terms up to order $\calO(g^2)$ are
\begin{equation}\label{pert.propagator}
\begin{split}
\deltabar(\vk_1+\vk_2) \: D_{i_1i_2}^{a_1a_2}(\vk_1) &=
D_0(1,2) \left[ 1-g^2 \: \perbraket{0}{1}{0}{1} \right] + \\
& + g^2 \bigg[ \perbra{0}{1} A(1) \, A(2) \perket{0}{1} +
\bra{0} A(1) \, A(2) \perket{0}{2} + \perbra{0}{2} A(1) \, A(2) \ket{0} \bigg]
\end{split}
\end{equation}
where $D_0$ is the tree-level propagator given in Eq.~(\ref{tree.level.gluon}).
As for the static ghost propagator, there are no terms of $\calO(g)$, since
the connected pieces of  $\bra{0} A \, A \perket{0}{1}$ are given by
the expectation value of five gauge field operators, which vanishes.

The normalisation factor [second term in the first line of Eq.~\eqref{pert.propagator}] 
cancels the disconnected piece of
the first term in the second line.
From Eq.~(\ref{pert.propagator}) it is also clear that the contributions to
$\perket{0}{2}$ with more than two-gluon states do not contribute
to $D(\vk)$ to the order considered.

The matrix elements in Eq.~\eqref{pert.propagator} can be straightforwardly
evaluated using the results of Sec.~\ref{sec:ham}. Moreover, the last
term in the second line of Eq.~\eqref{pert.propagator}
can be expressed in terms of the preceeding one since
\[
\perbra{0}{2} A(1) \, A(2) \ket{0} = 
\perbra{0}{2} A^\dagger(-1) \, A^\dagger(-2) \ket{0} =
\left( \bra{0} A(-1) \, A(-2) \perket{0}{2} \right)^* \: .
\]

\begin{figure}
\centering\includegraphics[width=.4\textwidth]{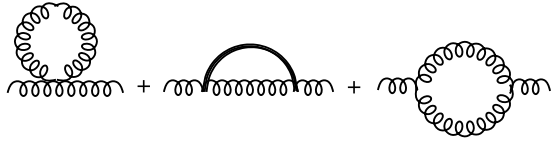}
\caption{\protect\label{fig.gluon} Diagrammatic representation of
Eq.~\eqref{res:1}. The curly and the full lines represent,
respectively, the static gluon and Coulomb propagators.}
\end{figure}

For the gluon form factor $D_A(\vk)$ defined by
\be\label{gluon.form.factor}
D_{ij}^{ab}(\vk) =: \delta^{ab} \, t_{ij}(\vk) \: \frac{D_A(\vk)}{2 |\vk|}
\ee
we get at one-loop level
\begin{align}
D_A(\vk) ={}& 1 - g^2 \mu^{3-d} \: \frac{N_c}{8 (d-1) \vk^2} \int \dbar{q} \:
\frac{d^2 - 3d + 3 - (\uvk\cdot\uvq)^2}{|\vq|} + \nonumber \\
& + g^2 \mu^{3-d} \: \frac{N_c}{4(d-1)} \int \dbar{q} \: \frac{t_{ij}(\vk) \, t_{ij}(\vq)}{(\vk-\vq)^2}
\left[ \frac{1}{|\vq|} - \frac{|\vq|}{\vk^2} \right] + \label{res:1} \\
&+ g^2 \mu^{3-d} \: \frac{N_c}{8(d-1)\vk^2} \int \!\! \dbar{q} \, \dbar{p} \,
\frac{T(\vp,\vk,\vq) (2\pi)^d \delta(\vp+\vk+\vq)}{(|\vp|+|\vk|+|\vq|) \, |\vp| \, |\vq|}
\left[ 1 + \frac{|\vk|}{|\vp|+|\vk|+|\vq|} \right] , \nonumber
\end{align}
with $\uvk=\vk/|\vk|$ and
\be\label{vertex}
T(\vk_1,\vk_2,\vk_3) = \tr|T(1,2,3)|^2 \: ,
\ee
where $T(1,2,3)$ is defined in Eq.~\eqref{3.gluon.vertex} and
the trace is taken in Lorentz space, yielding
\begin{align}
T(\vk,\vq,\vp) ={} & 2 (\vp^2 + \vq^2 + \vk^2) \big[ d - 2 - (\uvp \cdot \uvq) (\uvq \cdot \uvk) (\uvp \cdot \uvk) \big] + \nonumber \\
& + 2 (d-1) \big[ (\uvp \cdot \vq) (\uvp \cdot \vk) + (\uvq \cdot \vp) (\uvq \cdot \vk) + (\uvk \cdot \vq) (\uvk \cdot \vp)
- \vp \cdot \vq - \vq \cdot \vk - \vp \cdot \vk \big] + \nonumber \\
& - 2 (d-3) \big[ (\uvp\cdot\uvq)^2 (\vp^2+\vq^2) + (\uvq\cdot\uvk)^2 (\vq^2+\vk^2) + (\uvp\cdot\uvk)^2 (\vp^2+\vk^2) \big]
\end{align}
The diagrammatic representation of Eq.~\eqref{res:1} is shown
in Fig.~\ref{fig.gluon}.\footnote{Notice that these graphs are \emph{not} standard Feynman diagrams.}

The first integral in Eq.~\eqref{res:1} is a tadpole term, which vanishes
identically in dimensional regularisation. Moreover, in the last line
of Eq.~\eqref{res:1}  we can carry
out one of the momentum integrals due to the $\delta$-function, yielding
\begin{align}
D_A(\vk) = 1 &+ g^2 \, \mu^{3-d} \: \frac{N_c}{4(d-1)\vk^2} \int \dbar{q} \: \frac{d - 2 + (\uvk\cdot\uvq)^2}{(\vk-\vq)^2}
\: \frac{\vk^2-\vq^2}{|\vq|} + \nonumber \\
&+ g^2 \, \mu^{3-d} \: \frac{N_c}{2(d-1)\vk^2} \int \dbar{q} \,
\frac{\Sigma(\vk,\vq)}{|\vq| \, |\vk-\vq|} \:
\frac{2|\vk|+|\vq|+|\vk-\vq|}{\left( |\vk|+|\vq|+|\vk-\vq| \right)^2} \nonumber \\
=: 1 &+ I_c(\vk) + I_g(\vk) \label{res:2}
\end{align}
with
\begin{equation}\label{sigma}
\begin{split}
\Sigma(\vk,\vq) &= t_{il}(\vk) \, t_{jm}(\vq) \, t_{kn}(\vk-\vq)
\big[ \delta_{ij} k_k - \delta_{jk} q_i - \delta_{ik} k_j \big] 
\big[ \delta_{lm} k_n - \delta_{mn} q_l - \delta_{nl} k_m \big] \\
&= \left[ 1-(\uvk\cdot\uvq)^2 \right]
\left[ (d-1)(\vk^2 + \vq^2) + \frac{(d-2) \, \vk^2 \, \vq^2 + (\vk\cdot\vq)^2}{(\vk-\vq)^2} \right] .
\end{split}
\end{equation}

The first integral in Eq.~\eqref{res:2}, \textit{i.e.} the contribution
$I_c(\vk)$ from the Coulomb kernel, can be evaluated in $d=3-2
\varepsilon$
dimensions by means of standard techniques, yielding
\be
\label{Ic-dimreg}
I_c(\vk) = \frac{g^2 \, N_c}{(4\pi)^{2-\varepsilon}} \left[ \frac{4}{15} \left(
\frac{1}{\varepsilon} - \ln\frac{\vk^2}{\mu^2} - \gamma \right)
-\frac{8}{15} \: \ln 2 + \frac{188}{225} + \calO(\varepsilon) \right] .
\ee
Unfortunately, the gluon loop $I_g(\vk)$ [second integral
of Eq.~\eqref{res:2}] is highly non-trivial. It would probably
be possible to evaluate $I_g(\vk)$ using partial differential
equations techniques similar to the ones used in
Refs.~\cite{WatRei07a,WatRei07b}. Instead of using these
techniques, we will show that the integrals in Eq.~\eqref{res:2}
are the same as the ones treated in Refs.~\cite{WatRei07a,WatRei07b},
and we will use the result of those papers.

%%%%%%%%%%%%%%%%%%%%%%%%%%%%%%%%%%%%%%%%%%%%%%%%%%%%%%%%%%%

\subsection{\label{subsec:peter}Static gluon propagator from the Lagrangian approach}

In the Lagrangian-based functional integral approach to Yang--Mills theory in
Coulomb gauge considered in Refs.~\cite{WatRei07a,WatRei07b}, 
the full (energy-dependent) propagator has the form
\be\label{edepgp}
\langle A_i^a(p) A_j^b(k) \rangle = (2\pi)^{d+1} \, \delta(p+k) \, \delta^{ab} 
\, t_{ij}(\vk) \, W(k_4,\vk) \: ,
\ee
where $W(k_4,\vk)$ can be expressed (in Euclidean space) as
\be\label{fulltimeprop}
W(k_4,\vk) = \frac{D_{AA}(k_4,\vk)}{k_4^2+\vk^2} \: .
\ee
Here we have introduced the dressing function $D_{AA}(k_4,\vk)$, which measures
the deviation of the propagator from the tree-level form.
We are interested here in the \textit{static} or \textit{equal-time} propagator, the quantity considered in
the Hamiltonian approach, which is obtained from $W (k_4, \vk)$ by
integrating out the temporal component of the 4-momentum
\be\label{etgp}
W(\vk) = \int \frac{\d k_4}{2\pi} \: W(k_4,\vk) \: .
\ee
At tree-level (where $D_{AA}=1$) this yields
\be\label{e.t.prop.lag}
W_0(\vk) = \frac{1}{2 |\vk|} \: ,
\ee
which is precisely the static tree-level gluon propagator
of the Hamiltonian approach, see Eq.~(\ref{tree.level.gluon}). 
For sake of comparison with the Hamiltonian approach,
Eqs.~\eqref{gluon.form.factor} and \eqref{res:2}, we also
express the full \textit{equal-time} gluon propagator
\eqref{etgp} by a dressing function $\bar{D}_{AA}(\vk)$
\be\label{G73}
W(\vk) =: \bar{D}_{AA}(\vk) \, W_0(\vk) \stackrel{\eqref{e.t.prop.lag}}{=} \frac{\bar{D}_{AA}(\vk)}{2 |\vk|}
\ee
The two dressing functions (form factors) in \eqref{fulltimeprop} and
\eqref{G73} are related by
\be
\label{F9B-X1}
\bar{D}_{AA} (\vk) =  2 |\vk| \int \frac{\d k_4}{2 \pi} \frac{D_{AA} (k_4,
\vk)}{k^2_4 + \vk^2} \: .
\ee
With the dressing function of the energy-dependent propagator $D_{AA}(k_4, \vk)$
given in Refs.~\cite{WatRei07a,WatRei07b},
we can calculate the dressing function $\bar{D}_{AA} (\vk)$ of the \textit{equal-time}
propagator. We will now show that the so obtained \textit{equal-time} dressing function
of the gluon propagator Eq.~\eqref{F9B-X1}
coincides, at one-loop level,
with the gluon form factor of the Hamiltonian approach, defined in
Eq.~\eqref{gluon.form.factor}.

In Refs.~\cite{WatRei07a,WatRei07b} the dressing function $D_{AA}(k_4,\vk)$ is 
evaluated at one-loop level, with the result%
\footnote{The results of \cite{WatRei07a} and \cite{WatRei07b}, being evaluated
in the, respectively, first and second order formalism, are at first sight
not identical. However, it is not difficult to show that they indeed agree.}
\begin{equation}\label{peter}
\begin{split}
D_{AA}(k_4,\vk) = 1 &+
g^2 \, \mu^{3-d} \: \frac{2 N_c}{(d-1)} \int 
\frac{\d[d]q \, \d q_4}{(2\pi)^{d+1}}
\frac{\Sigma(\vk,\vq)}{k^2\,  q^2 \, (k-q)^2} + \\
&+ g^2 \, \mu^{3-d} \: \frac{N_c}{(d-1)} \int \frac{\d[d]q \, \d q_4}{(2\pi)^{d+1}}
\frac{t_{ij}(\vk)t_{ij}(\vq)}{k^2 \, q^2 \, (\vk-\vq)^2} \, (k_4^2-\vq^2) \: .
\end{split}
\end{equation}
Here $\Sigma$ is the kernel obtained by contracting the three-gluon vertex,
defined in Eq.~\eqref{sigma}. Inserting Eq.~(\ref{peter}) in
Eq.~\eqref{F9B-X1} and also performing the loop integration over $q_4$ one
obtains for the \textit{equal-time} dressing function $\bar{D}_{AA} (\vk)$ (\ref{F9B-X1})
precisely the form factor $D_A (\vk)$ of the static gluon propagator of the
Hamiltonian approach, Eq.~(\ref{res:2}).
We have thus shown that the Lagrangian-based functional integral approach yields
the same \textit{equal-time} gluon propagator as the time-independent Hamiltonian approach,
at least to the order considered.
We have checked that this equivalence does also hold for the $\langle \Pi \Pi
\rangle$ correlator, but it does \emph{not} hold for the $\langle A\Pi \rangle$ 
correlator. The reason is that in the time-dependent Lagrangian approach the
$\langle A\Pi \rangle$  correlator is odd under time reversal and thus the
corresponding \textit{equal-time} correlator vanishes, while in the Hamiltonian approach
the static $\langle A\Pi \rangle$   correlator is constrained by the canonical
commutation relation not to vanish.

In Refs.~\cite{WatRei07a,WatRei07b}, the loop corrections to the gluon dressing function
Eq.~(\ref{ghost-dimreg}) were calculated in dimensional regularisation.
With $d = 3 - 2\varepsilon$ the result reads
\begin{equation}\label{peter.result}
\begin{split}
D_{AA}(k_4,\vk) = 1 & + \frac{g^2 \, N_c}{(4\pi)^{2-\varepsilon}} 
\bigg\{ \bigg[ \frac{1}{\varepsilon} -\gamma - 
\ln\frac{\vk^2}{\mu^2} \bigg] - \ln(1+z) - \frac{64}{9} +3z + \\
& + \frac{1}{4} \: f(z) \bigg[\frac{1}{z}-1-11z-3z^2 \bigg]+
g(z) \bigg[-\frac{1}{2z}+\frac{14}{3}-\frac{3}{2} \: z \bigg] + \calO(\varepsilon) \bigg\} \, ,
\end{split}
\end{equation}
where $z=k_4^2/\vk^2$ and the functions $f(z)$, $g(z)$ are defined by
\begin{subequations}\label{G78}
\begin{align}
f(z) &= 4 \, \ln2 \, \frac{\arctan\sqrt{z}}{\sqrt{z}} - \int_0^1 \d t \: \frac{\ln(1+z\,t)}{\sqrt{t} \, (1+z\,t)} \\
g(z) &= 2 \, \ln2 - \ln(1+z)
\end{align}
\end{subequations}

Using this result, after integrating Eq.~(\ref{peter.result}) over $k_4$
with the appropriate tree-level factor, see Eq.~\eqref{F9B-X1},
we find for the static gluon form factor (\ref{gluon.form.factor})
\be\label{gluon-dimreg}
D_A(\vk) =1 + g^2 \; \frac{N_c}{(4\pi)^{2-\varepsilon}}
 \left[ \left( \frac{1}{\varepsilon} 
 - \ln\frac{\vk^2}{\mu^2} \right) + \ldots \right]
\ee
where the ellipsis contains the finite constant terms.

Finally, we remark that the ghost propagator in the Lagrangian-based
functional integral approach \cite{WatRei07a,WatRei07b} defined there as
the correlator $\langle c \bar{c} \rangle$ for explicitly introduced ghost
and antighost fields, is given by a function $W_c(k_4,\vk)$ in analogy with
Eq.~\eqref{edepgp} for the gluon propagator, which is \emph{independent}
of the temporal component $k_4$, $W_c(k_4,\vk) = W_c(\vk)$. It is easily seen,
by integrating over the temporal component of the loop momentum, that 
$W_c(\vk)$ coincides with our result \eqref{875-37} for
$G(\vk)=D_c(\vk)/\vk^2$.

%%%%%%%%%%%%%%%%%%%%%%%%%%%%%%%%%%%%%%%%%%%%%%%%%%%%%%%%%%%
%%%%%%%%%%%%%%%%%%%%%%%%%%%%%%%%%%%%%%%%%%%%%%%%%%%%%%%%%%%

\section{The ghost-gluon vertex and the $\beta$-function}

In the Hamiltonian approach the ghost-gluon vertex is given by \cite{PhDSchleifenbaum}
\be\label{ghost-gluon-vertex}
\begin{split}
&\bra{\Omega} A_i^a(\vx) \, G^{bc}[A](\vy_1,\vy_2) \ket{\Omega} =: \\
&\int \d[d]z_1 \, \d[d]z_2 \, \d[d]z_3 \,
D_{ij}(\vx,\vz_1) \, G(\vy_1,\vz_2) \, G(\vy_2,\vz_3) \, \Gamma^{abc}_{i}(\vz_1;\vz_2,\vz_3) \, ,
\end{split}
\ee
where $D$ is the full gluon propagator of Sec.~\ref{sec:gluon} and $G$
the static ghost propagator as defined in Sec.~\ref{sec:ghost}. Eq.~\eqref{ghost-gluon-vertex}
tells us to calculate the vacuum expectation value $\langle A G[A] \rangle$
and then ``cut off'' the external legs, in order to extract the irreducible
component. In Eq.~\eqref{ghost-gluon-vertex} we already used the fact
that ghost and gluon propagators are colour diagonal to every order in
perturbation theory.

\begin{figure}
\centering\includegraphics[width=.3\linewidth]{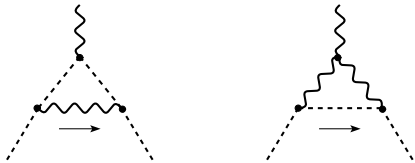}
\caption{\label{fig:ggv} Diagrammatic representation of the one-loop
correction to the ghost-gluon vertex. The left and right picture correspond
to, respectively, the first and second integral in Eq.~\eqref{ggv}. The arrow
shows the flow of the loop momentum $\vl$.}
\end{figure}

The lowest-order contribution to Eq.~\eqref{ghost-gluon-vertex} is
the bare vertex $\Gamma^{(0)}$, given in momentum space by
\be\label{bare-ggv}
\Gamma^{(0)abc}_{i}(\vk;\vp,\vq) = - i \, g \, f^{abc} p_i \, ,
\ee
(all momenta defined as incoming). With arguments similar to the ones
we used in the evaluation of the ghost and gluon propagators, it is not
difficult to show that the terms contributing to the next non-vanishing
order are
\[
\bra{0} A G_2[A] \perket{0}{1} + \perbra{0}{1} A G_2[A] \ket{0} +
\bra{0} A G_3[A] \ket{0}_\mathrm{1PI} \, ,
\]
where the subscript \mbox{1PI} means that only the irreducible
terms have to be considered. The evaluation of these matrix
elements is straightforward and yields
\be\label{ggv}
\begin{split}
\Gamma^{(2)abc}_i(\vk;\vp,\vq) ={}& - i \, g^3 \, f^{abc} \frac{N_c}{4} \:
\bigg\{ \int \dbar{\ell} \:
\frac{(\ell-p)_i \, p_j \, q_m \, t_{jm}(\vl)}{|\vl| (\vp-\vl)^2 (\vq+\vl)^2} + \\
&+ \int \dbar{\ell} \:
\frac{p_j \, q_m}{\vl^2 |\vp-\vl| |\vq+\vl|}
\frac{T_{ijm}(\vk,\vp-\vl,\vq+\vl)}{|\vk|+|\vp-\vl|+|\vq+\vl|}
\bigg\}
\end{split}
\ee
where $T$ is the Lorentz structure of the three-gluon vertex \eqref{3.gluon.vertex}. The diagrammatic
representation of Eq.~\eqref{ggv} is given in Fig.~\ref{fig:ggv}.
The integrals in Eq.~\eqref{ggv} are UV finite in $d=3$ spatial dimensions,
hence they are independent of the scale when evaluated at a symmetry
point (after factorising the momentum that carries the Lorentz index $i$).
Then $g^2 \, D_A \, D_c^2$ is a renormalisation group invariant,
at least to the present order. From Eqs.~\eqref{ghost-dimreg} and
\eqref{gluon-dimreg} we obtain in Coulomb gauge
\begin{align*}
D_A(\vk) D_c^2(\vk) &= \left[ 1 + g^2 \; \frac{N_c}{(4\pi)^{2-\varepsilon}} 
\; \frac{1}{\varepsilon} + \ldots \right]
\left[ 1 + g^2 \; \frac{N_c}{(4\pi)^{2-\varepsilon}} 
\; \frac{4}{3} \; \frac{1}{\varepsilon}  + \ldots \right]^2 = \\
&= 1 + g^2 \; \frac{N_c}{(4\pi)^{2-\varepsilon}} \; \frac{11}{3} \; 
\frac{1}{\varepsilon} + \ldots
\end{align*}
Consequently, $g^2$ must have a $1/\varepsilon$ pole with coefficient $(-11/3) N_c$, and since,
at the one-loop level, the coefficient
of the $1/\varepsilon$ pole of $g^2$ is the (gauge invariant)
first coefficient $\beta_0$ of the $\beta$-function,
\be
\beta(g) = \frac{\partial g}{\partial\ln\mu} = \frac{1}{(4\pi)^2} \, \beta_0 \, g^3 + \calO(g^5) \, ,
\ee
we find
\be\label{beta0}
\beta_0 = - \frac{11}{3} \: N_c \: ,
\ee
which is the correct value.

%%%%%%%%%%%%%%%%%%%%%%%%%%%%%%%%%%%%%%%%%%%%%%%%%%%%%%%%%%%
%%%%%%%%%%%%%%%%%%%%%%%%%%%%%%%%%%%%%%%%%%%%%%%%%%%%%%%%%%%

\section{\label{section:coulombpotential}The potential for static sources}

Until now, we have considered pure Yang--Mills theory without external
colour charges. If static external sources are included, the
Hamilton operator becomes
\begin{equation}
\begin{split}
\widetilde{H} &= \widetilde{H}_\mathrm{YM} +
\frac{g^2}{2} \int \d[d]x \, \d[d]y \, \rho_m^a(\vx) F^{ab}[A](\vx,\vy) \rho_m^b(\vy) + \\
& + \frac{g^2}{2} \int \d[d]x \, \d[d]y \left[
\rho_m^a(\vx) F^{ab}[A](\vx,\vy) \Hat{A}_i^{bc}(\vy) \widetilde{\Pi}_i^c(\vy) 
+  \Hat{A}_i^{ac}(\vx) \widetilde{\Pi}_i^{c\dagger}(\vx) F^{ab}[A](\vx,\vy) \rho_m^b(\vy) \right] ,
\end{split}
\end{equation}
where $\widetilde{H}_\mathrm{YM}$ is the Hamiltonian \eqref{h-YM}
of the Yang--Mills sector considered in the previous sections and
$\rho^a_m(\vx)$ is the external charge density. Since
\[
\widetilde{\Pi}^a_i(\vx) = \Pi^a_i(\vx) + \frac{i}{2} \frac{\delta\ln\calJ}{\delta A_i^a(\vx)} \, , \qquad
\widetilde{\Pi}^{a\dagger}_i(\vx) = \Pi^a_i(\vx) - \frac{i}{2} \frac{\delta\ln\calJ}{\delta A_i^a(\vx)} \, ,
\]
the $\delta\ln\calJ/\delta A$ contributions cancel in the terms
linear in the external charge density and we can omit the
tildes there, obtaining
\begin{equation}\label{hamiltonian}
\begin{split}
\widetilde{H} = \widetilde{H}_\mathrm{YM} &+
\frac{g^2}{2} \int \d[d]x \, \d[d]y \: \rho_m^a(\vx) F^{ab}[A](\vx,\vy) \rho_m^b(\vy) + \\
&+ \frac{g^2}{2} \int \d[d]x \, \d[d]y \: \anticomm{\rho_m^a(\vx) F^{ab}[A](\vx,\vy)}{\Hat{A}^{bc}_i(\vy) \, \Pi_i^c(\vy)} .
\end{split}
\end{equation}
($\{,\}$ denotes the anticommutator.)
We are interested here in a perturbative calculation
of the static potential, \textit{i.e.} the potential
between static charges. For this purpose, we will
follow the approach of Refs.~\cite{Gri77,Dre81,book_Lee},
treating the external sources as perturbations to the pure
Yang--Mills sector. Suppose we know the exact spectrum
of $\widetilde{H}_\mathrm{YM}$, \textit{i.e.} its eigenstates
$\ket{\Phi_N}$ and the corresponding eigenvalues $E_{\Phi_N}$.
Second-order perturbation theory in the external sources
then yields for the potential
\begin{equation}\label{potential}
V^{ab}(\vx,\vy) = \bra{\Phi_0} F^{ab}[A](\vx,\vy) \ket{\Phi_0} - \frac{g^2}{2} \sum_{N\neq0}
\frac{\bra{\Phi_0} K^{a}(\vx) \ket{\Phi_N} \bra{\Phi_N} K^{b}(\vy) \ket{\Phi_0}}{E_{\Phi_N}-E_{\Phi_0}} \: ,
\end{equation}
where we have introduced the quantity
\be
K^{a}(\vx) := \int \d[d]z \: \anticomm{F^{ab}[A](\vx,\vz)}{\Hat{A}^{bc}_i(\vz) \Pi^c_i(\vz)} .
\ee

Since we are interested in the potential to order $\calO(g^2)$,
we can replace all quantities in the second term in
Eq.~\eqref{potential} by their unperturbed expressions. This yields
\begin{equation}\label{pot.lin}
- \frac{g^2}{2} \sum_{N\neq0}
\frac{\bra{\Phi_0} K^{a}(\vx) \ket{\Phi_N} \bra{\Phi_N} K^{b}(\vy) \ket{\Phi_0}}{E_{\Phi_N}-E_{\Phi_0}} =
- g^2 \sum_{1,2} \frac{|\bra{0} F_0 \Hat{A} \Pi \ket{g_1,g_2}|^2}{|\vk_1| + |\vk_2|} + \calO(g^4) ,
\end{equation}
where the unperturbed part of $F^{ab}[A](\vx,\vy)$, Eq.~\eqref{coulomb.operator}, in $d=3$ spatial
dimensions
\begin{equation}
F_0^{ab}(\vx,\vy) = \delta^{ab} \, \left[ (-\partial^2)^{-1} \right]_{\vx,\vy}
=\frac{\delta^{ab}}{4\pi|\vx-\vy|},
\end{equation}
is the familiar Coulomb interaction.

In the first term of Eq.~\eqref{potential}
(the so-called colour Coulomb potential)
we use the perturbative expansion
of the vacuum wave functional $|\Omega\rangle$, Eq.~\eqref{full.vacuum}, and
also expand the Coulomb kernel $F^{ab}[A](\vx,\vy)$, Eq.~\eqref{coulomb.operator},
in powers of $g$. Thereby many terms can be shown to vanish or cancel using similar
arguments as in the evaluation of the static ghost propagator. In the end one
obtains for the Coulomb potential
\begin{equation}\label{pot.quad}
\bra{\Omega} F^{ab}[A](\vx,\vy) \ket{\Omega} =
F_0^{ab}(\vx,\vy) + g^2 \, \bra{0} F^{ab}_2[A](\vx,\vy) \ket{0} + \calO(g^4) \; ,
\end{equation}
where $F^{ab}_2[A](\vx,\vy) = 3 G^{ab}_2[A](\vx,\vy)$
is the $\calO(g^2)$ part of the Coulomb kernel $F[A]$, Eq.~\eqref{coulomb.operator}.

\begin{figure}
\centering\includegraphics[width=.3\linewidth]{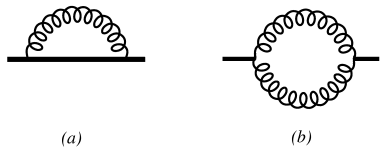}
\caption{\protect\label{fig.coulomb}Diagrams contributing to the static
potential in second order perturbation theory. \textit{(a)} and \textit{(b)}
represent, repectively, the second term of Eq.~\eqref{pot.quad} and Eq.~\eqref{pot.lin}.}
\end{figure}

For the potential dressing function $v(\vk)$ defined by
\be\label{potentialformfactor}
V^{ab}(\vx,\vy) = \int \dbar{k} \: e^{i \vk \cdot (\vx-\vy)} \frac{\delta^{ab} \, v(\vk)}{\vk^2},
\ee
we obtain the following expression
\begin{equation}\label{potential-1loop}
\begin{split}
v(\vk) = 1 &+ g^2 \mu^{3-d} \frac{3 N_c}{2 \vk^2} \int \dbar{q} \: \frac{k_i k_j t_{ij}(\vq)}{(\vk-\vq)^2 |\vq|} + \\
&- g^2 \mu^{3-d} \frac{N_c}{2 \vk^2} \int \dbar{q} \: \frac{t_{ij}(\vq) \, t_{ij}(\vk-\vq)}{|\vq|}
\frac{|\vk-\vq|-|\vq|}{|\vk-\vq|+|\vq|} \\
= 1 &+ v_\mathrm{I}(\vk) + v_\mathrm{II}(\vk) \: .
\end{split}
\end{equation}
The diagrammatic representation of this equation is given
in Fig.~\ref{fig.coulomb}. The first integral of Eq.~\eqref{potential-1loop} comes from the
expectation value of $F_2$ [second term in 
Eq.~\eqref{pot.quad}], while the second integral arises from
the second-order term [Eq.~\eqref{pot.lin}].

While the first integral in Eq.~\eqref{potential-1loop} is standard,
the second one is of the same type we encountered in the calculation
of the static gluon propagator. Again, we will use the results of the
Lagrangian-based functional integral approach. As argued by Zwanziger
\cite{Zwa98}, the potential between static colour charges
should be related to the correlation function of the $A_0$ field.
For the corresponding dressing function
$D_\sigma$ defined by
\be
\langle A_0^a(p) A_0^b(k) \rangle = (2\pi)^{d+1} \, \delta(p+k) \, \delta^{ab} \,
\frac{D_\sigma(k_4,\vk)}{\vk^2} \: ,
\ee
the result at one-loop order reads \cite{WatRei07a}
\be\label{a0a0corr}
\begin{split}
D_\sigma(k_4,\vk) = 1 &+ g^2 \mu^{3-d} \, \frac{3N_c}{2\vk^2} \int \dfr[d]{q}
\frac{k_i k_j t_{ij}(\vq)}{q^2 (\vk-\vq)^2} + \\
&+ g^2 \mu^{3-d} \frac{N_c}{2 \, \vk^2} \int \dfr[d]{q}
\frac{(\vk^2 - 2 \vk\cdot\vq) \: t_{ij}(\vq) \, t_{ij}(\vk-\vq)}{|\vk-\vq| \, [k_4^2 + (|\vk-\vq|+|\vq|)^2 ]} \: ,
\end{split}
\ee
where we have already performed the integration over the temporal
component $q_4$ of the loop momentum. Noticing that
\be
\vk^2 - 2 \vk\cdot\vq = (\vk-\vq)^2-\vq^2 = (|\vk-\vq|+|\vq|) (|\vk-\vq|-|\vq|) \, ,
\ee
we see that at one-loop level
\be
v(\vk) = D_\sigma(k_4=0,\vk) \, .
\ee
Taking then the result for $D_\sigma$ obtained in \cite{WatRei07a}
\be\label{peter:sigma}
\begin{split}
D_\sigma(k_4,\vk)= 1 + \frac{g^2 N_c}{(4\pi)^{2-\varepsilon}} \bigg\{ {}&
\frac{11}{3} \bigg[\frac{1}{\varepsilon} - \gamma - \ln\frac{\vk^2}{\mu^2} - \ln(1+z) \bigg] + \frac{31}{9} - 6z + \\
& + (3z-1) g(z) + \frac12 (1+z)(1+3z) f(z) + \calO(\varepsilon) \bigg\} \, ,
\end{split}
\ee
where $z=k_4^2/\vk^2$ and the functions $f(z)$, $g(z)$ are defined in
Eqs.~\eqref{G78}, we get for the dressing function of the static potential
\be\label{potential-1loop-1}
v(\vk) = 1 + g^2 \, \frac{N_c}{(4\pi)^{2-\varepsilon}}
\left\{ \frac{11}{3} \left[ \frac{1}{\varepsilon} - \gamma - \ln\frac{\vk^2}{\mu^2} \right]
+ \frac{31}{9} + \calO(\varepsilon) \right\} .
\ee
A few remarks are in order here. The equivalence between the static
potential evaluated in the Hamiltonian approach and the propagator of
the $A_0$ field in the Lagrangian-based formalism is far from trivial,
since in the Hamiltonian approach the Weyl gauge $A^a_0=0$ is also
imposed. Furthermore, note that the Hamiltonian static potential is
not obtained as the \textit{equal-time} component, but rather as the
integral over the relative time of the $\langle A_0 A_0 \rangle$
correlator. Such an integral captures the instantaneous interaction
through the first correction term in Eq.~\eqref{a0a0corr} as much as
the retarded interaction via vacuum polarisation given by the second
correction term there.

Given the fact that the physical potential is $g^2 V^{ab}$, with
$V^{ab}$ given by Eq.~\eqref{potentialformfactor}, $v(\vk)$ in Eq.~\eqref{potential-1loop}
represents the form factor of the running coupling. From Eq.~\eqref{potential-1loop-1}
we then find $\beta_0 = - 11 N_c/3$, which is again the correct
coefficient. Since this
agrees with $D_A(\vp)D_c^2(\vp)$, we find the same running
coupling from the ghost-gluon vertex and the Coulomb
potential, at least to the order considered.

%%%%%%%%%%%%%%%%%%%%%%%%%%%%%%%%%%%%%%%%%%%%%%%%%%%%%%%%%%%
%%%%%%%%%%%%%%%%%%%%%%%%%%%%%%%%%%%%%%%%%%%%%%%%%%%%%%%%%%%

\section{Relation with the variational approach}

In the variational approach considered in Refs.~\cite{FeuRei04,EppReiSch07},
a generalisation of Eq.~\eqref{pert.vacuum} was taken for the vacuum functional,
\be\label{non.pert.vacuum}
\langle {A}|{\omega} \rangle = \calN \exp \left\{ - \frac{1}{2} \int \dfr[3]{k} \:
A_i^a(\vk) \, t_{ij}(\vk) \, \omega(\vk) \, A_j^a(-\vk) \right\} ,
\ee
and the kernel $\omega(\vk)$ was determined by minimisation
of the vacuum energy density. 
In this non-perturbative case a complete basis can also be defined
by Eqs.~\eqref{G28} and \eqref{G31} with $|\vk|$ replaced by $\omega(\vk)$.
Considering the three-gluon vertex as a perturbation on top of the non-perturbative
vacuum \eqref{non.pert.vacuum}, the energy functional gets an additional contribution
\be\label{en3}
\Delta E[\omega] =  - \frac{g^2}{3!} \sum_{1,2,3}
\frac{|\bra{\omega} \widetilde{H}_1\ket{g_1,g_2,g_3}|^2}{\omega(\vk_1)+\omega(\vk_2)+\omega(\vk_3)} \: .
\ee
Functional differentiation of this expression with respect to $\omega(\vk)$ yields an
additional term to the gap equation
\begin{equation}\label{gapequation}
\begin{split}
\omega^2(\vk) &= \vk^2 + \chi^2(\vk) + I_\omega^0 + I_\omega(\vk) + \\
& - g^2 \; \frac{N_c}{8} \int \dfr[3]q \, \dfr[3]p \;
\frac{T(\vk,\vq,\vp) \; (2\pi)^3 \delta(\vp+\vq+\vk)}{[\omega(\vk)+\omega(\vq)+\omega(\vp)] \, \omega(\vp) \, \omega(\vq)} 
\frac{2 \omega(\vk) + \omega(\vq)+\omega(\vp)}{\omega(\vk)+\omega(\vq)+\omega(\vp)} \: ,
\end{split}
\end{equation}
where $T(\vk,\vq,\vp)$ is given in Eq.~\eqref{vertex} and the integral
terms $\chi(\vk)$, $I_\omega^0$, and $I_\omega(\vk)$ are defined in
Ref.~\cite{Epp+07}. For $\vk\to\infty$ this additional term reduces to the
integral $I_g(\vk)$, Eq.~\eqref{res:2}, found in perturbation theory, showing
that Eq.~\eqref{gapequation} does indeed provide the right contribution to
the variational form of the gap equation to produce the correct UV asymptotic
behaviour, at the same time leaving the ghost dominated infrared sector
untouched. The numerical solution of this modified gap equation is in progress.

%%%%%%%%%%%%%%%%%%%%%%%%%%%%%%%%%%%%%%%%%%%%%%%%%%%%%%%%%%%

\section{Summary and Conclusions}

In this paper we have studied the Hamiltonian approach to Yang--Mills theory
in Coulomb gauge in Rayleigh--Schr\"odinger perturbation theory. The static gluon and
ghost propagators as well as the potential between static colour sources
have been calculated to one-loop order using dimensional regularization.
The one-loop $\beta$-function was calculated from the ghost-gluon
vertex as well as from the static potential. In both cases the result known from
covariant perturbation theory was reproduced. The unperturbed basis constructed
from the eigenstates of the unperturbed Hamiltonian, which up to the colour index
of the gauge field coincides with the Hamiltonian for QED (without fermions),
was generalized to multiple quasi-gluon states on top of the non-perturbative
vacuum wave functional used in the variational approach \cite{FeuRei04,EppReiSch07}.
Treating the three-gluon vertex, which is not captured by the variational
wave functional used so far, as a perturbation on top of the non-perturbative
vacuum, a modified gap equation was derived, which yields the correct
(perturbative) ultraviolet asymptotics for the static propagators
known from perturbation theory while, at the same time, leaving the
(non-perturbative) infrared behaviour of the propagator unchanged. The multiple quasi-gluon basis
constructed on top of the non-perturbative vacuum wave functional will also serve
as a basis for calculating the partition function of Yang--Mills theory in the
Hamiltonian approach in Coulomb gauge and investigating the deconfinement phase
transition at finite temperatures.

%%%%%%%%%%%%%%%%%%%%%%%%%%%%%%%%%%%%%%%%%%%%%%%%%%%%%%%%%%%

\begin{acknowledgments}
The authors would like to thank Wolfgang Schleifenbaum
and Peter Watson for valuable discussions.
This work was supported by the Deutsche Forschungsgemeinschaft (DFG) under
contracts No.~Re856/6-1 and No.~Re856/6-2, the
Cusanuswerk--Bisch\"ofliche Studienf\"orderung,
the Deutscher Akademischer Austauschdienst (DAAD), CIC-UMSNH,
and Conacyt project No.~46513-F.
A. W. is grateful to the Institute for Theoretical Physics at the University
of T\"ubingen for the warm hospitality extended to him during a two-months 
stay in the summer of 2008.
\end{acknowledgments}

%%%%%%%%%%%%%%%%%%%%%%%%%%%%%%%%%%%%%%%%%%%%%%%%%%%%%%%%%%%

\end{document}